\documentclass[page-classic]{revtex4}

\usepackage{graphicx}
\usepackage{dcolumn}
\usepackage{bm}
\usepackage{epsfig}
\usepackage{amsmath}
\usepackage{mathrsfs}
\usepackage[dvips]{color}

\begin{document}

\title{The $\Sigma\Sigma$ interactions in finite-density QCD sum rules}
\author{
       P.\ Li$^1$\footnote{E-mail:
       lipengburnlife@mail.nankai.edu.cn},
       X.\ H.\ Zhong$^2$\footnote{E-mail: zhongxh@ihep.ac.cn},
       L.\ Li$^1$\footnote{E-mail: lilei@nankai.edu.cn}, and
       P.\ Z. Ning$^1$\footnote{E-mail: ningpz@nankai.edu.cn}
       }

\affiliation{
  $^1$ Department of Physics, Nankai University, Tianjin 300071, China\\
  $^2$ Institute of High Energy Physics,Chinese Academy of Sciences, Beijing 100039, China
  }

\begin{abstract}
The properties of $\Sigma$-hyperons in pure $\Sigma$ matter are
studied with the finite-density quantum chromo-dynamics sum rule
(QCDSR) approach. The $\Sigma\Sigma$ nuclear potential $U_\Sigma$ is
most likely strongly attractive, it could be about $-50$ MeV or even
more attractive at normal nuclear density. If this prediction is the
case, the interactions between $\Sigma$-hyperons should play crucial
roles in the strange nuclear matter, when there are multi-$\Sigma$
hyperons. The bound state of double-$\Sigma$ maybe exist.
\end{abstract}



\maketitle


\section{INTRODUCTION}

The baryon-baryon interactions are basic issues in nuclear
physics. In the past years, there has been a lot of work on the
nucleon-nucleon interactions ($NN$), for there exists much
experimental information from nucleon-nucleon scattering. However,
the interactions of hyperon-nucleon ($YN$) are much less known
than these of $NN$ due to the difficulties in performing
scattering experiments with the unstable hyperons. In these $YN$
interactions, only the $\Lambda N$ interaction is known for us by
studying the hyper-nuclei. The hyperon-hyperon ($YY$) interactions
are the least known ones in baryon-baryon interactions for the
very scarce information in experiments. More studies of $YN$, $YY$
interactions are needed not only by the development of the nuclear
physics, but also by the application in the other fields, such as
in astrophysics.

There have been some typical models for the study of $YN$ and $YY$
interactions. Such as the $SU(6)$ quark model
\cite{Fujiwara:2006yh,Fujiwara:2001xt,Nakamoto:1997gh}, the
$SU(3)$ chiral quark model \cite{Zhang:1997ny,Shen:1999pf}, the
chiral effective field theory
\cite{Polinder:2007mp,Polinder:2006zh,Polinder:2006tb}, the
lattice QCD \cite{Savage:2006pn}, the chiral unitary approach
\cite{Sasaki:2006cx}, the meson-exchange model
\cite{Haidenbauer:2005zh,Stoks:1999bz} and the QCDSR
\cite{a1,b2,b3,b4,b5}.

In this work, we will study the $\Sigma\Sigma$ interactions with
finite-density QCDSR, which had been developed in the series
papers \cite{aaa,a1,a3,a4,b1,b2,b3,b4,b5,b6}. With this approach,
the properties of nucleons, $\Lambda$- and $\Sigma$-hyperons in
nucleonic nuclear matter have been reasonably described. Based on
the sum rules for $\Sigma N$ interactions \cite{b5}, the
$\Sigma\Sigma$ interactions can be described by substituting the
in-medium condensates in nucleonic nuclear matter with pure
$\Sigma$ matter, this approach has been adopted in our previous
work, in which the $\Lambda$$\Lambda$ interactions were discussed
\cite{a2}.

The finite-density QCDSR approach focuses on a correlation
function of interpolating fields, made up of quark fields, which
carry the quantum numbers of the hadron of interest. Unlike usual
ones, the correlation function is evaluated in the ground state of
the nuclear matter rather than in vacuum. The correlation function
can represent in a simple phenomenological ansatz for these
spectral densities on  the one hand. On the other hand, the
correlation function can be evaluated  at large space-like momenta
using an operator product expansion (OPE). Finally, one can deduce
the sum-rules by equating these two different representations
using appropriately weighted integrals. The baryon self-energies
in medium matter can be related  to QCD Lagrangian parameters and
finite-density condensates.

For simplicity, only the leading order of the in-medium
condensates are taken into account in this work, which is a
reasonable approximation at low nuclear densities
\cite{Plohl:2007dp,Kaiser:2007nv}. In the OPE for $\Sigma$
correlation function, we consider all condensates up to dimension
4, and the terms up to the first order in the strange quark mass
$m_s$. In addition, the contributions from the dimension-6
four-quark condensate are included for their importance. And the
leading order in-medium gluon condensates,
$\langle\overline{q}q\rangle_\rho$,
$\langle\overline{s}s\rangle_\rho$, \mbox{$\langle
\frac{\alpha_s}{\pi} [(u'\cdot
G)^2+(u'\cdot\tilde{G})^2]\rangle_\rho$},
$\langle\frac{\alpha_s}{\pi}G^2\rangle_\rho$, $\langle q^\dagger i
D_0 q\rangle_\rho$ and $\langle s^\dagger i D_0 s\rangle_\rho$ are
derived from the chiral perturbation theory (ChPT). Following the
Ref.~\cite{a1}, to deal with  the determined scalar-scalar
four-quark condensate $\langle\overline{q}q\rangle_\rho^2 $, we
introduce a arbitrary parameter $f$ to describe its density
dependence.

The paper is organized as follows. In the subsequent section, the
sum rules and the condensates are given.  The calculations and
analysis are presented in Sec. \ref{result}.  Section
\ref{summary} is a summary.

\section{THE METHOD}\label{su}

\subsection{ QCDSR for $\Sigma$ hyperons in $\Sigma$ matter }

The finite-density QCDSR approach has been well devolved in the
series lectures \cite{aaa,a1,a3,a4,b1,b2,b3,b4,b5}. As done in
\cite{a2}, we can easily extent the $\Sigma$ sum rules in nuclear
matter \cite{a1} to describe the $\Sigma \Sigma$ interactions in
the pure $\Sigma$ matter by changing the quark and gluon
condensates in nuclear matter to those in pure $\Sigma$ matter
(the sum rules are listed in appendix A). Using the obtained sum
rules, the baryon scalar self-energy $\Sigma_s$ and the vector
self-energy $\Sigma_v$ and the effective mass $M^*_\Sigma$ can be
related to the in-medium quark and gluon condensates at
finite-density. Then, the $\Sigma\Sigma$ nuclear potential
$U_\Sigma$ can be valued by the formula $U_\Sigma = \Sigma_s +
\Sigma_v$. The essential quark and gluon condensates are
calculated in the subsequent section.

\subsection{In-medium condensates}

To obtain the predictions for the $\Sigma\Sigma$ interactions in
pure $\Sigma$ matter from the sum rules described above, we need
to know the condensates in pure $\Sigma$ matter. The first order
of the condensates in the nuclear matter can be written as
\begin{eqnarray}\label{eqn 8}
\langle \hat{\mathcal{O}}\rangle_\rho=\langle
\hat{\mathcal{O}}\rangle_0 + \langle \hat{\mathcal{O}}\rangle_{
\Sigma} \rho+\ldots,
\end{eqnarray}
where the ellipsis denote the corrections of higher order density,
 and $\langle \hat{\mathcal{O}}\rangle_{ \Sigma}$ is the spin
averaged $\Sigma$ matrix element.

As we know, in the QCD Hamiltonian density $\mathcal{H}_{QCD}$,
chiral symmetry is explicitly broken by the current quark mass
terms. Neglecting the isospin breaking effects, one has the
Hamiltonian \cite{b1}:
\begin{eqnarray}
\mathcal{H}_{mass} \equiv 2 m_q \overline{q}q+m_s
\overline{s}s+\ldots,
\end{eqnarray}
where $m_s$ and $m_q$ are the strange and light $u,d$ current
quark masses, respectively; $q$ and $s $ stand for the $u,d$ quark
and strange quark fields, respectively. Taking the Hamiltonian
$\mathcal{H}_{mass}$ as a function of $m_q$, in the
Hellmann-Feyman theorem, one obtains
\begin{eqnarray}
\lefteqn{2 m_q \langle \Psi(m_q)|\int
\mathrm{d}x^3\overline{q}q|\Psi(m_q)\rangle} {}\nonumber,\\
&&{}=m_q\frac{\mathrm{d}}{\mathrm{d}m_q}\langle \Psi(m_q)|\int
\mathrm{d}x^3\mathcal{H}_{mass}|\Psi(m_q)\rangle.
\end{eqnarray}
In the above equation, we consider the cases of
$|\Psi(m_q)\rangle=|vac\rangle$ and
$|\Psi(m_q)\rangle=|\rho\rangle$, where
$|\Psi(m_q)\rangle=|\rho\rangle$ denotes the ground state of
$\Sigma$ matter with $\Sigma$ density $\rho$ and
$|\Psi(m_q)\rangle=|vac\rangle$ denotes the vacuum state. Taking
the difference of these two cases, and taking into account the
uniformity of the system yields
\begin{eqnarray}\label{a1}
2
m_q(\langle\overline{q}q\rangle_\rho-\langle\overline{q}q\rangle_0)=m_q
\frac{\mathrm{d}\mathcal{E}}{\mathrm{d} m_q},
\end{eqnarray}
where $\mathcal{E}$ is the energy density of the $\Sigma$ matter,
which is given by
\begin{eqnarray}\label{eqn 5}
\mathcal{E}=M_{\Sigma} \rho+\delta \mathcal{E},
\end{eqnarray}
where $\delta \mathcal{E}$ is of the higher order term. Recently,
the in-medium condensates are studied in Refs.
\cite{Plohl:2007dp,Kaiser:2007nv}, from their analysis it is found
that the contributions of the higher order term $\delta
\mathcal{E}$ to the in-medium condensates are small at low density
$\rho\leq \rho_0$. Thus, the contributions of the higher order
term $\delta \mathcal{E}$ are neglected in the calculations. In
the chiral perturbation theory (see the Appendix B of \cite{a2} ),
the $\Sigma$ mass is given by
\begin{eqnarray}\label{eqn 17}
M_\Sigma=\lefteqn{M_N+4(b_D-b_F)B_0 m_s{}}\nonumber \\
&& \quad -4(b_D-b_F)B_0 m_q,
\end{eqnarray}
where $b_D$, $b_F$ and $B_0$ are real parameters in the chiral
Lagrangian, which can be seen in many references for example
\cite{a2}. Then, from the Eq. (\ref{a1}),  we obtain
\begin{eqnarray}\label{m_q}
\langle\overline{q}q\rangle_\rho=\langle\overline{q}q\rangle_0+\frac{1}{2
m_q} \big[\sigma_{\pi N}-4 m_q(b_D-b_F)B_0 \big]\rho,
\end{eqnarray}
where $\sigma_{\pi N}$ is the $\pi N$ sigma term, which is given as
\begin{eqnarray}
\sigma_{\pi N}=m_q \frac{\mathrm{d}M_N}{\mathrm{d}m_q}.
\end{eqnarray}

Following the steps above and those in Ref.~\cite{a2}, the other
dimension 3 and 4 quark and gluon condensates are easily obtained.
The results are
\begin{eqnarray}
&& \langle q^\dagger q \rangle_\rho = \langle u^\dagger
u\rangle_\rho =\langle d^\dagger d  \rangle_\rho=\langle s^\dagger
s
\rangle_\rho=\rho, \\
&&
\langle\overline{s}s\rangle_\rho=\langle\overline{s}s\rangle_0+\frac{1}{m_s}
\big[S-4 m_s(b_F-b_D)B_0 \big]\rho , \\
&&
 \langle q^\dagger i D_0 q\rangle_\rho =
 \frac{m_q}{4}\langle\overline{q}q\rangle_\rho
+\frac{3}{8}M_\Sigma[A^u_2 (\mu^2)+A^d_2 (\mu^2)] \rho, \nonumber\\
&& \\
&&
 \langle s^\dagger i D_0 s\rangle_\rho =
 \frac{m_s}{4}\langle\overline{s}s\rangle_\rho
+\frac{3}{4}M_\Sigma[A^s_2 (\mu^2)] \rho ,\\
&& \langle \frac{\alpha_s}{\pi}G^2 \rangle_\rho=\langle
\frac{\alpha_s}{\pi}G^2 \rangle_0 -\frac{8}{9} \{ M_\Sigma - [
\sigma_{\pi N} + S +\mathcal{K}]\}\rho, \nonumber \\
&& \\
&& \bigg\langle \frac{\alpha_s}{\pi} \bigg[(u'\cdot
G)^2+(u'\cdot\tilde{G})^2\bigg]\bigg\rangle_\rho=-\frac{3}{2\pi}M_\Sigma
\mathcal{C}(\mu^2) \rho, \nonumber \\
\end{eqnarray}
where $S=m_s \frac{\mathrm{d}M_N}{\mathrm{d}m_s}=\frac{1}{2}
\frac{m_s}{m_q} \sigma_{\pi N} \, y$ \cite{b1} is the strangeness
content of nucleon with a dimensionless quantity $y\equiv
\langle\overline{s}s\rangle_N / \langle\overline{q}q\rangle_N$ ,
the moments of parton distribution functions $A^Q_2 (\mu^2)$ in
$\Sigma$ hyperon matter are $A^u_2+A^d_2\simeq A^s_2 \simeq 0.3$
\cite{a2}, and the value of the
$\mathcal{C}(\mu^2)=\alpha_s(\mu^2) A^g_2(\mu^2)$ is about $0.22$
\cite{b2}, and the parameters $\mathcal{K}$ express as
$\mathcal{K}=4(m_s- m_q )(b_D-b_F)B_0$. The other parameters, such
as the vacuum condensates and the current quark masses, are
adopted the same as those in our previous work \cite{a2}.


Finally, the in-medium four-quark condensate, $
\langle\overline{q}q\rangle_\rho^2$, should be considered justly,
because they are numerically important in the finite density sum
rules. As pointed out in Refs. \cite{b2,b3}, the in-medium
four-quark condensates in the $\Sigma$ sum rules are their
factorized forms, which may not be justified in nuclear matter
because the four-quark condensates are sensitive to the nuclear
density, one might suspect that this is an artifact of the
factorization. Thus, as done in Refs. \cite{b2,b3} we choose to
parameterize the scalar-scalar four quark condensates so that they
interpolate between their factorized form in free space and their
factorized form in $\Sigma$ matter. That is, in the calculations
we need replace $\langle\overline{q}q\rangle_\rho^2$ in
Eqs.~(\ref{A1}, \ref{A2}) by modified form
$\langle\widetilde{\overline{q}q}\rangle_\rho^2$:
\begin{eqnarray}
\langle\widetilde{\overline{q}q}\rangle_\rho^2=(1-f)\langle\overline{q}q\rangle_0^2
+ f \langle\overline{q}q\rangle_\rho^2,
\end{eqnarray}
where $f$ is the real parameter. The predictions in
Refs.~\cite{b2,b3,b4,a2,a1} suggest that the four-quark
condensate, $\langle\overline{q}q\rangle_\rho^2$, should depend
weakly on the nuclear density. That is, the artificial parameter
$f$ is most possibly in the range of $0\leq f \leq 0.5$.

\section{CALCULATIONS AND ANALYSIS}\label{result}

In the calculations, to quantify the fit of the left- and right-
sides of the $\Sigma$ sum rules, we use the logarithmic measure
\cite{a1,a2,b2,b3,b4,b5,b6}
\begin{widetext}
\begin{eqnarray}\label{log}
\delta(M^2) &=& \mathrm{ln}\Bigg[ \frac{\max \Big\{
{\lambda^{*2}e^{-(E^2_q-\mathbf{q}^2)/M^2},
\Pi'_s(M^2)/M^{*}_\Sigma, \Pi'_q(M^2), \Pi'_u(M^2)/\Sigma_v}
\Big\} }{\min \Big \{ {\lambda^{*2}e^{-(E^2_q-\mathbf{q}^2)/M^2},
\Pi'_s(M^2)/M^{*}_\Sigma, \Pi'_q(M^2), \Pi'_u(M^2)/\Sigma_v}
\Big\} } \Bigg].
\end{eqnarray}
\end{widetext}
Here $\Pi'_s(M^2)$, $\Pi'_q(M^2)$ and $\Pi'_u(M^2)$ denote the
right-hand sides of the Eqs.(\ref{A1}--\ref{A3}), respectively. In
principle, this three terms are equal to
$\lambda^{*2}e^{-(E^2_q-\mathbf{q}^2)/M^2}$.   The predictions for
$\lambda^{*2}$, $s^{*}_0$,  $M^{*}_\Sigma$,  $\Sigma_v$ are
obtained by minimizing the measure $\delta$. In the zero-density
density, we can obtain the $\Sigma$ mass in vacuum applying the
same procedure to the sum rules.

\subsection{Borel mass}

Firstly, we should choose a proper Borel mass $M^2$ in the
calculation. In principle the predictions should be independent of
the Borel mass $M^2$. However, in practice one has to truncate the
operator product expansion and use a simple phenomenological
ansatz for the spectral density, which cause the sum rules to
overlap only in some limited range of $M^2$. The previous studies
for the octet baryons show that the sum rules do not provide a
particulary convincing plateau. Nevertheless, we can assume that
the sum rules actually has a region of overlap, although it is
imperfect.  In order to compensate for at least some of the
limitations of the truncated sum rules, we normalize the
finite-density predictions for all self-energies to the
zero-density prediction for the mass. In Refs. \cite{a1,b2,b3,a2},
the optimization region of $M^2$ is suggested as $0.8\leq M^2\leq
1.4$ GeV$^2$, thus, in this work we choose the proper Borel mass
$M^2$ around this region.

\begin{figure}[ht]
\center
 \epsfig{file=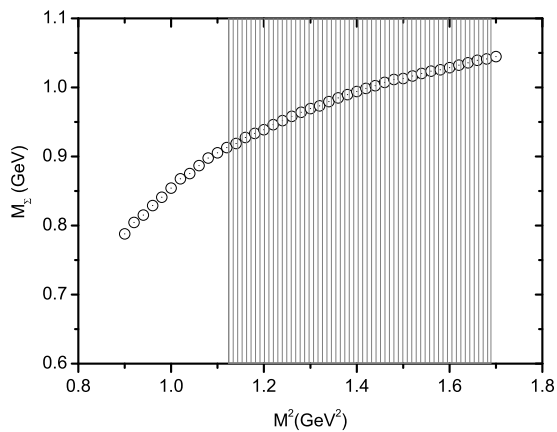,width=8.2cm} \caption{ $\Sigma$ mass in vacuum as a
 function of the auxiliary parameter $M^2$}
 \label{pic-vac}
\end{figure}

\begin{figure}[ht]
\center \epsfig{file=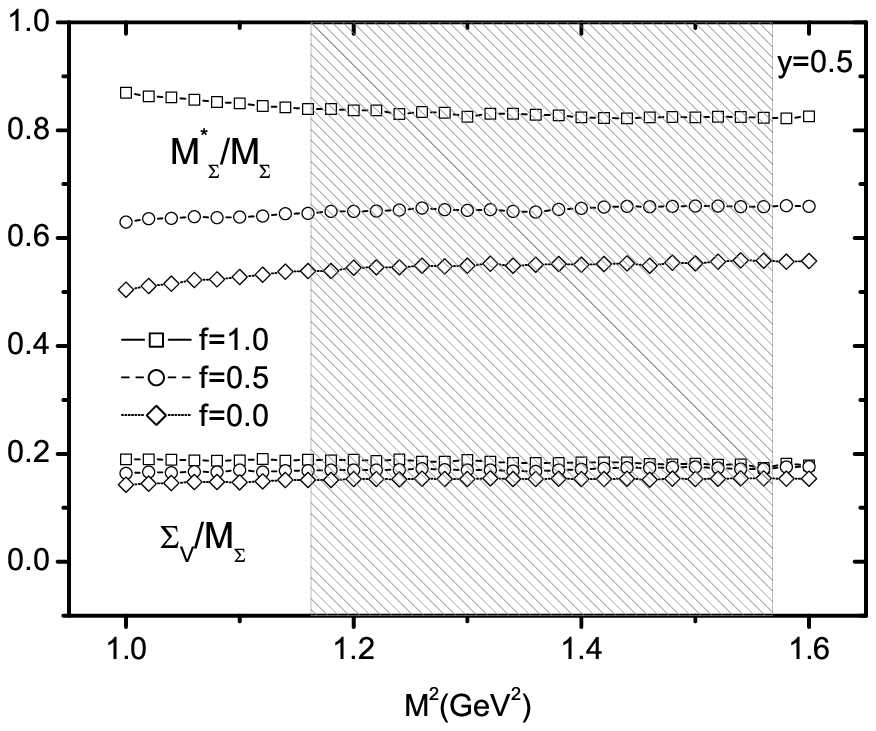, width=8.2cm} \caption{
$M^*_\Sigma/M_\Sigma$ and $\Sigma_V/M_\Sigma$ as functions of Borel
mass $M^2$, where y = 0.5, $\sigma_{\pi N}=56$ MeV and
$\rho=\rho_0=(110~\mathrm{MeV})^3$. The three curves correspond to
$f=0.0$ (diamond), $f=0.5$ (circle), and $f=1.0$ (square),
respectively.} \label{pic-M20.5}
\end{figure}

To find an optimization region for $M^2$ (in this region the
predictions should be less sensitive to $M^2$ than those in other
regions), we plot the $\Sigma$ masses in vacuum and in nuclear
medium as a function of Borel mass $M^2$ in its possible range
$0.9\leq M^2\leq 1.7 ~\mathrm{GeV^2}$ in Fig.~\ref{pic-vac} and
Fig.~\ref{pic-M20.5}, respectively. It is found that by
normalizing the finite-density predictions in the calculation, a
good plateau appears in the range of $1.1\leq M^2\leq
1.6~\mathrm{GeV^2}$. This optimal Borel mass predicted by us
consists with the previous predictions in Refs.
\cite{a1,b2,b3,a2}. In our later calculations, we choose the
medium value $M^2=1.4~\mathrm{GeV^2}$.

\subsection{Sensitivity to the $f$ and $|q|$}

\begin{figure}[ht]
\center
 \epsfig{file=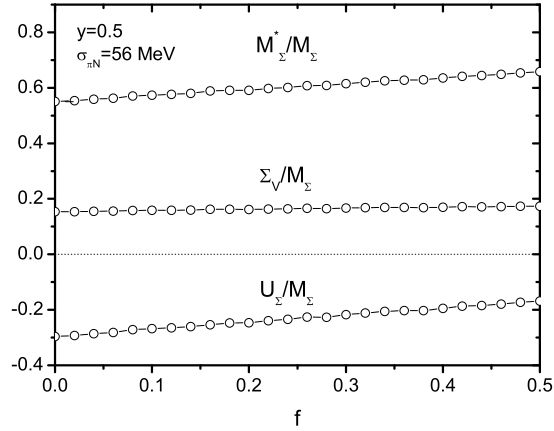, width=8.2cm} \caption{The $M^*_\Sigma/M_\Sigma$,
 $\Sigma_V/M_\Sigma$ and $U_\Sigma/M_\Sigma$ as functions of four-quark condensate parameter $f$.
 The other input
parameters are the same as in Fig.~\ref{pic-M20.5}.}
 \label{picf0.5}
\end{figure}

\begin{figure}[ht]
\center
 \epsfig{file=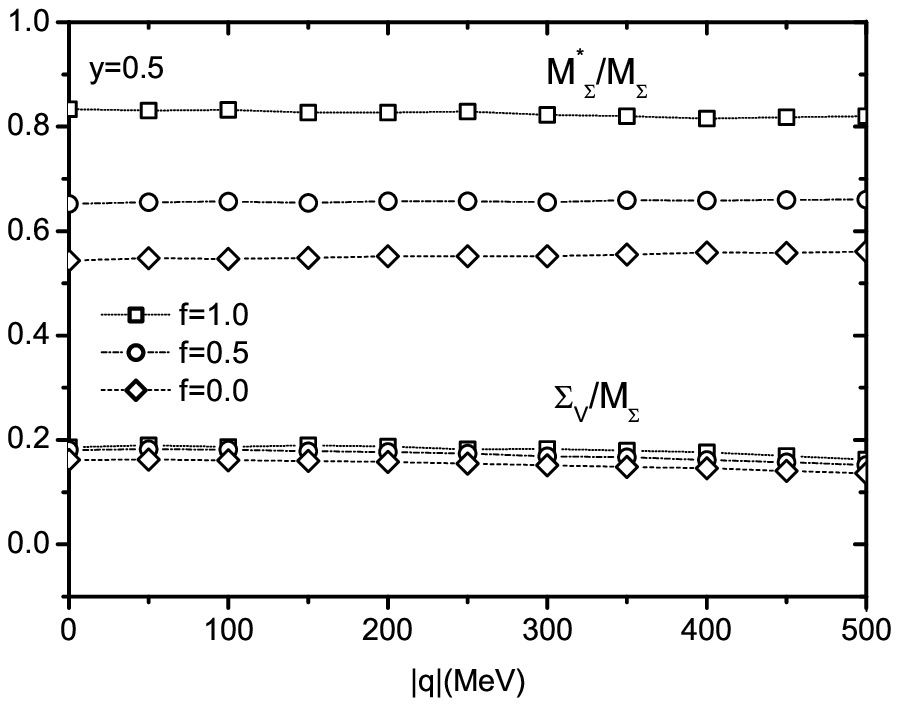, width=8.2cm} \caption{$M^*_\Sigma/M_\Sigma$ and
 $\Sigma_V/M_\Sigma$ as functions of three momentum $|q|$. The other input
parameters are the same as in Fig.~\ref{pic-M20.5}. }
 \label{picq0.5}
\end{figure}

Then, the sensitivity of the predictions to the $f$  is
illustrated in Fig.~\ref{picf0.5}, where $\sigma_{\pi N}$, $y$ and
$|q|$ are fixed at 56 MeV, 0.5 and 270 MeV, respectively, as done
in \cite{a2}. The Fig.~\ref{picq0.5} is about the optimum results
as a function of the momentum $|q|$ at the normal nuclear density
$\rho=\rho_0=(110 ~\mathrm{MeV})^3$ with three different values of
$f$. From the figures, it is seen that the predictions for
$M^*_\Sigma/M_\Sigma$ and $U_\Sigma / M_\Sigma$ are sensitive to
$f$ (i.e. the four-quark condensate) but slightly dependent on
$|q|$, they monotonously increase with the increment of the $f$.
The $U_\Sigma/M_\Sigma$ is insensitive to both the $f$ and $|q|$.

\subsection{Sensitivity to $\sigma_{\pi N}$ and $y$}

\begin{figure}[ht]
\center \epsfig{file=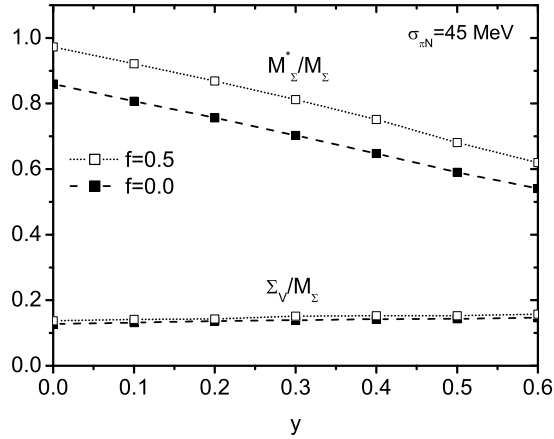, width=8.2cm}
\caption{$M^*_\Sigma/M_\Sigma$ and $\Sigma_V/M_\Sigma$  as
functions of $y$ with $\sigma_{\pi N}=45$ MeV. The other input
parameters are the same as in Fig.~\ref{pic-M20.5}.} \label{pic-y}
\end{figure}

\begin{figure}[ht]
\center \epsfig{file=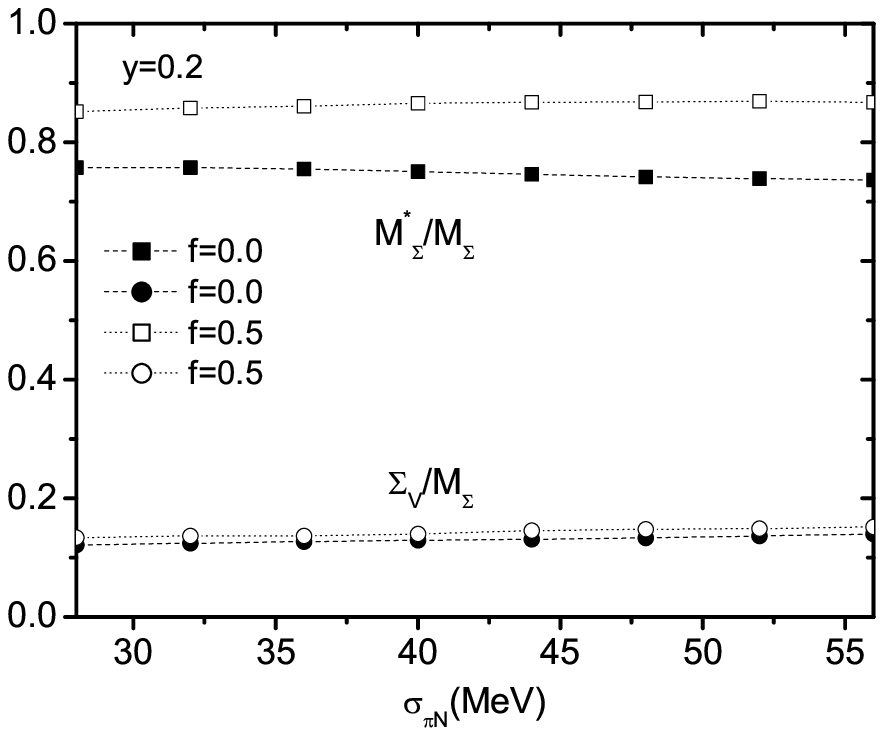, width=8.2cm} \caption{
$M^*_\Sigma/M_\Sigma$ and $\Sigma_V/M_\Sigma$ as functions of the
$\pi N$ sigma term $\sigma_{\pi N}$ with $y=0.2$. The other input
parameters are the same as in Fig.~\ref{pic-M20.5}.}
\label{pic-sigma}
\end{figure}

There are large uncertainties of the $\pi N$ sigma term
$\sigma_{\pi N}$ and the strangeness content of the nucleon $y$.
The recent determinations suggest large values for $\sigma_{\pi
N}=64\pm 8$ MeV, and hence a large strangeness content of the
nucleon ,i.e., $y=0.5$ are obtained. The $\Sigma N$ sum rule study
suggests large strangeness content $y=0.5$, which is also in
agreement with our recent predictions $y=0.5$ and $\sigma_{\pi
N}=$56 MeV in the study of the $\Lambda\Lambda$ interaction with
QCDSR. While the usual adopted values of $\sigma_{\pi N}$ and $y$
is $\sigma_{\pi N}=45$ MeV and $y\simeq 0.2$. To study the effect
of the parameters $y$ and $\sigma_{\pi N}$, we plot
$M^*_\Sigma/M_\Sigma$ and $\Sigma_V/M_\Sigma$ as functions of $y$
and $\sigma_{\pi N}$, respectively, in Fig.~\ref{pic-y} and
Fig.~\ref{pic-sigma}.

From the two figures, it can be seen that $\Sigma_V/M_\Sigma$ is
insensitive to both y and $\sigma_{\pi N}$. However, the scalar
self-energy is sensitive to the strange quark content $y$, and
slightly depends on the $\sigma_{\pi N}$. $M^*_\Sigma/M_\Sigma$
increases with the increment of the $f$, however, decreases with
the increment of the $y$.

\subsection{The predictions versus density}

Finally, to study the in-medium properties of the
$\Sigma$-hyperon, the effective mass $M^*_\Sigma/M_\Sigma$,  the
vector self-energy $\Sigma_v/M_\Sigma$ and the potential
$U_\Sigma/M_\Sigma$ as functions of densities $\rho$ are plotted.
For the uncertainties of the $\sigma_{\pi N}$ and $y$, two sets of
the $\sigma_{\pi N}$ and $y$ are adopted in this work. One set is
the new determinations $\sigma_{\pi N}=56$ MeV and $y=0.5$; and
the other set is the usual values $\sigma_{\pi N}=45$ MeV and
$y=0.2$.

\begin{figure}[ht]
\center
 \epsfig{file=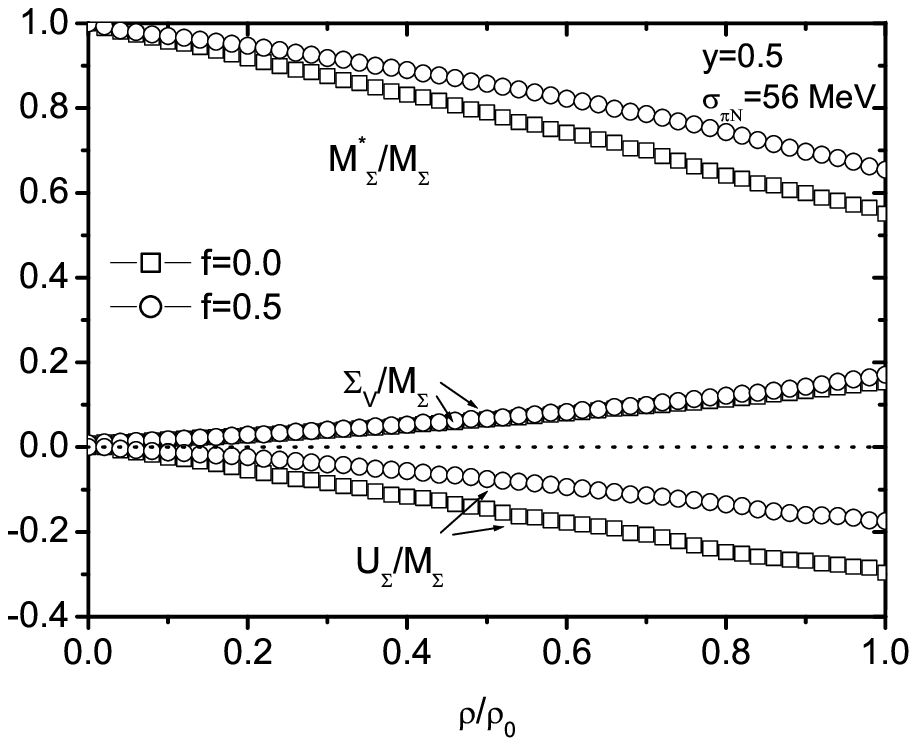, width=8.2cm} \epsfig{file=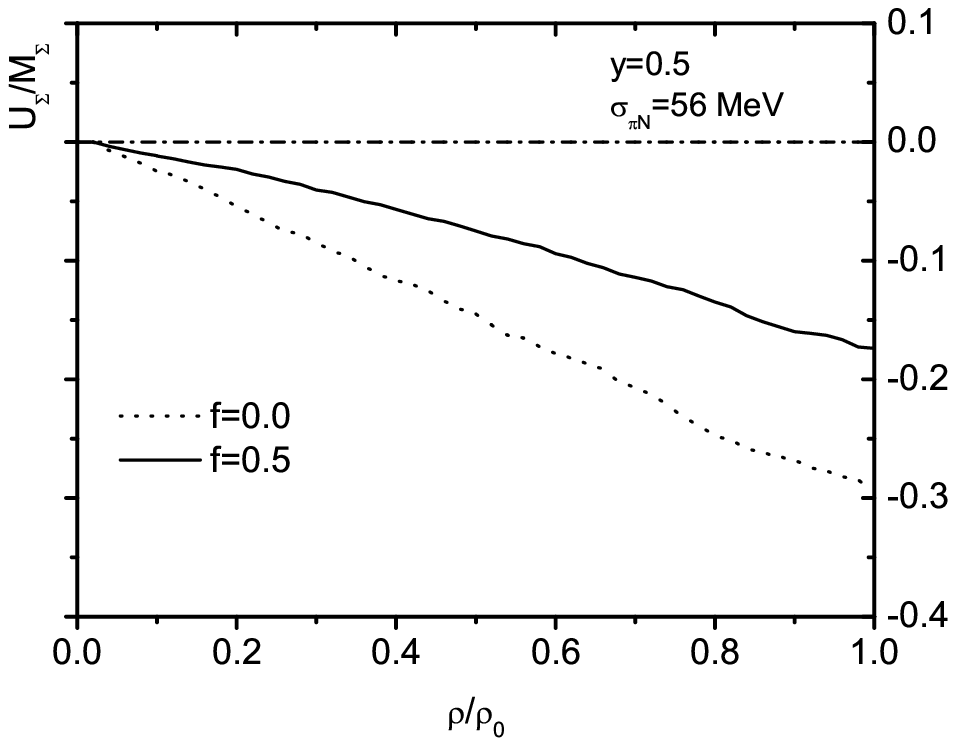, width=8.2cm}
 \caption{$M^*_\Sigma/M_\Sigma$, $\Sigma_V/M_\Sigma$ and $U_\Sigma/M_\Sigma$
 as functions of the $\Sigma$ density $\rho$ with $f=0.0$ (square) and
 $f=0.5$ (circle), respectively, where $y=0.5, \sigma_{\pi N}=56$ MeV. The other input parameters are the
 same as in the Fig.~\ref{pic-M20.5}.}
 \label{aa}
\end{figure}

\begin{figure}[ht]
\center
 \epsfig{file=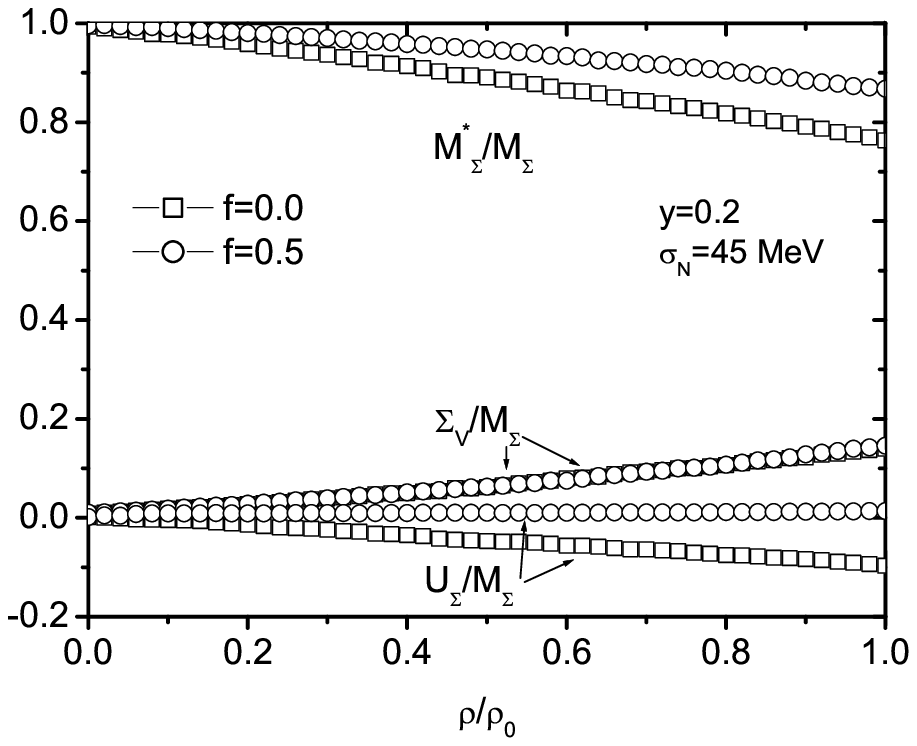, width=8.2cm}
 \epsfig{file=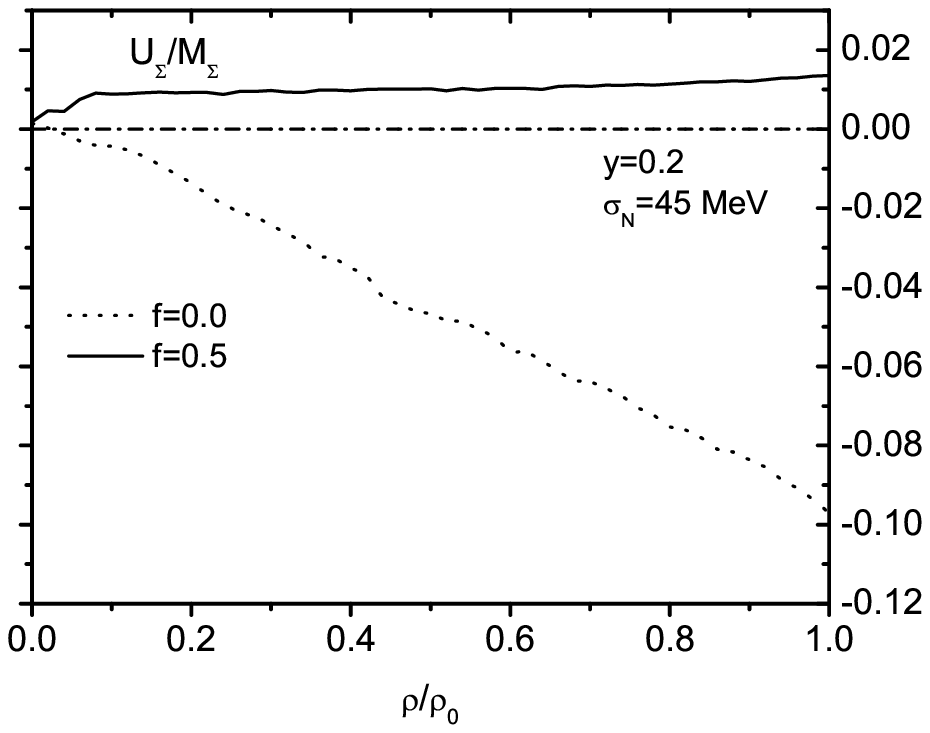, width=8.2cm} \caption{$M^*_\Sigma/M_\Sigma$, $\Sigma_V/M_\Sigma$ and $U_\Sigma/M_\Sigma$
 as functions of the $\Sigma$ density $\rho$ with $f=0.0$ (square) and
 $f=0.5$ (circle), respectively, where $y=0.2, \sigma_{\pi N}=45$ MeV. The other input parameters are the
 same as in the Fig.~\ref{pic-M20.5}.}
 \label{bb}
\end{figure}

From the figures \ref{aa} and \ref{bb}, we see the effective mass
$M^*_\Sigma/M_\Sigma$ decreases, whereas the vector self-energy
$\Sigma_v/M_\Sigma$ increases monotonously with the increment of
the $\Sigma$ density. The differences of the effective mass
$M^*_\Sigma/M_\Sigma$ between the parameter $f=0.0$ and $f=0.5$
are more and more obvious with the increment of the density
$\rho$, while the vector self-energy $\Sigma_v/M_\Sigma$ is
insensitive to the parameter $f$ at different densities.

From the figure, we also find that the potential $U_\Sigma$ has
strong parameter dependence in the whole density region $0 \leq
\rho \leq \rho_0$. The differences between the two sets $f=0.0$
and $f=0.5$ become more and more obvious with the increment of the
density.

When we set $\sigma_{\pi N}=56$ MeV, $y=0.5$, and
$|q|=270~\mathrm{MeV}$, the effective mass and vector self-energy
at $\rho_0$ are
\begin{eqnarray}
M^*_\Sigma \simeq (0.551-0.658) M_\Sigma, \\
\Sigma_v \simeq (0.153-0.173) M_\Sigma,
\end{eqnarray}
in the range $0\leq f \leq 0.5$ (see Fig. \ref{aa}); and the
potential $U_\Sigma$ is strongly attractive, the strength
increases monotonously with the increment of the $\Sigma$ density
$\rho$. At $\rho = \rho_0$, the potential can reach to
\begin{eqnarray}
U_\Sigma \simeq  -(296\sim 174)\  \mathrm{MeV},
\end{eqnarray}
which is much stronger than the nuclear potential of nucleon at
normal density. Similarly, the attractive $\Sigma\Sigma$ potential
is also predicted in \cite{Stoks:1999bz,Nakamoto:1997gh}, which is
even stronger than the $NN$ one.

If we set $\sigma_{\pi N}=45$ MeV, $y=0.2$ (see Fig. \ref{bb}), it
is  seen that the effective mass $M^*_\Sigma/M_\Sigma$, the vector
self-energy $\Sigma_v/M_\Sigma$ at $\rho_0$ are
\begin{eqnarray}
M^*_\Sigma \simeq (0.763\sim 0.863) M_\Sigma, \\
\Sigma_v \simeq (0.140 \sim 0.146) M_\Sigma,
\end{eqnarray}
and the potential is
\begin{eqnarray}
U_\Sigma \simeq -(50\pm 70) \ \  \mathrm{MeV}.
\end{eqnarray}
In this case, the medium value of the potential
$U_\Sigma\simeq-50$ MeV is also strongly attractive. Comparing
with the predictions of the two parameter sets, we find the vector
self-energies of them are almost equal, however, the nuclear
potential with $y=0.2, \sigma_{\pi N}= 45$ MeV are much weaker
than that with $y=0.5, \sigma_{\pi N}= 56$ MeV.

Although the sum-rule predictions for the scalar self-energy are
quite sensitive to the four-quark condensates in nuclear medium and
parameter $y$, according to the analysis of the four-quark
condensates in the series papers, we could predict that the
$\Sigma\Sigma$ potential is most likely strongly attractive. This
potential is much stronger than the $\Lambda\Lambda$ potential
\cite{a2} in the same conditions. Thus, when we deal with the
strange nuclear matter, if many $\Sigma$ hyperons appear, the
interactions between $\Sigma$ hyperons should play crucial roles.
According to our predictions, the bound state of double-$\Sigma$
maybe exist.

\section{Summary}\label{summary}

In this paper, the  $\Sigma\Sigma$ interactions are analyzed
carefully with the finite-density QCDSR approach. The sum-rule
analysis indicates that the vector self-energy $\Sigma_v$ is
insensitive to the sum rule parameters. However, the potential
$U_\Sigma$ and the scalar self-energy $\Sigma_s$ have strong
parameter dependence, especially, they are very sensitive to the
four quark condensates. Although the predictions strongly depend
on the undetermined parameters $f$ and $y$, it can predict that
the $\Sigma\Sigma$ potential $U_\Sigma$ is most likely strongly
attractive, which could be $-50$ MeV or even more attractive at
normal nuclear density. If this prediction is the case, the
interactions between $\Sigma$ hyperons should play crucial roles
in the strange nuclear matter, when there are multi-$\Sigma$
hyperons.  The bound state of double-$\Sigma$ maybe exist.

This is a preliminary attempt to study the $\Sigma\Sigma$
interactions in finite $\Sigma$ density. More studies are needed
to describe the details of the potential. The four quark
condensate in medium should be studied further also.

\acknowledgements

This work is supported, in part, by the Natural Science Foundation
of China (grant 10575054, 10775145), China Postdoctoral Science
Foundation, and K. C. Wong Education Foundation, Hong Kong.

\appendix

\section{THE $\Sigma$ SUM RULES}

The sum rules for the $\Sigma$ hyperon in the nuclear matter had
been deduced by Xuemin Jin and Marina Nielsen\cite{a1}, which are
given by







\begin{widetext}

\begin{eqnarray}\label{A1}
\lambda^{*2}_\Sigma M^*_\Sigma e^{-(E^2_q-\mathbf{q}^2)/M^2}
\lefteqn{ =\frac{m_s}{16 \pi^4} M^6 E_2 L^{- 8/9} - \frac{M^4}{4
\pi^2}E_1 \langle\overline{s}s\rangle_\rho
 + \frac{m_s}{2\pi^2}
\overline{E}_q M^2 E_0 (\langle q^\dagger q \rangle_\rho - \langle
s^\dagger s \rangle_\rho) L^{-8/9}}
\nonumber \\
& & { } +\frac{4 m_s}{3 \pi^2}\textbf{q}^2 \langle q^\dagger i D_0
q\rangle_\rho L^{-8/9}
 + \frac{4 m_s}{3 \pi^2}
\langle\overline{q}q\rangle_\rho^2  -\frac{4}{3} \overline{E}_q
\langle\overline{s}s\rangle_\rho \langle q^\dagger q \rangle_\rho,
\end{eqnarray}
\begin{eqnarray}\label{A2}
\lambda^{*2}_\Sigma e^{-(E^2_q-\mathbf{q}^2)/M^2} \lefteqn{
=\frac{M^6}{32 \pi^4} E_2 L^{-4/9}  +\frac{M^2}{144\pi^2} (E_0 - 4
\frac{\textbf{q}^2}{M^2})   \times \bigg\langle
\frac{\alpha_s}{\pi} \bigg[(u'\cdot
G)^2+(u'\cdot\tilde{G})^2\bigg]\bigg\rangle_\rho L^{-4/9} }
\nonumber \\
&& { } + \frac{m_s}{18 \pi^2} M^2 (5 E_0 - 2
\frac{\textbf{q}^2}{M^2}) \langle\overline{s}s\rangle_\rho
L^{-4/9}  + \frac{M^2}{32 \pi^2}
  \times \langle\frac{\alpha_s}{\pi}G^2\rangle_\rho E_0 L^{-4/9}
  \nonumber \\
&& { } - \frac{4 M^2}{9 \pi^2}( E_0 - \frac{\textbf{q}^2}{M^2})
\times \langle q^\dagger i D_0 q\rangle_\rho  L^{-4/9}   -
\frac{M^2}{9 \pi^2} ( E_0 - 4 \frac{\textbf{q}^2}{M^2})  \langle
s^\dagger i D_0 s\rangle_\rho L^{-4/9} \nonumber \\
&& { } + \frac{\overline{E}_q}{6 \pi^2} M^2 E_0 (\langle q^\dagger
q \rangle_\rho + \langle s^\dagger s\rangle_\rho) L^{-4/9}  +
\frac{4}{3}\langle q^\dagger q \rangle_\rho \langle s^\dagger
s\rangle_\rho L^{-4/9} +
\frac{2}{3}\langle\overline{q}q\rangle_\rho^2 L^{4/9},
\end{eqnarray}
\begin{eqnarray}\label{A3}
\lambda^{*2}_\Sigma \Sigma_v e^{-(E^2_q-\mathbf{q}^2)/M^2}
\lefteqn{= \frac{1}{12 \pi^2} M^4 E_1 (7 \langle q^\dagger q
\rangle_\rho+\langle s^\dagger s\rangle_\rho) L^{-4/9}   -
\frac{\overline{E}_q}{9 \pi^2} M^2 E_0 ( m_s
\langle\overline{s}s\rangle_\rho }
\nonumber \\
&& { } - 16 \langle q^\dagger i D_0 q\rangle_\rho  - 4 \langle
s^\dagger i D_0 s\rangle_\rho) L^{-4/9} +\langle s^\dagger s
\rangle_\rho)L^{-4/9}
\nonumber \\
&& {} - \frac{\overline{E}_q}{36 \pi^2} M^2 E_0 \bigg\langle
\frac{\alpha_s}{\pi} \bigg[(u'\cdot
G)^2+(u'\cdot\tilde{G})^2\bigg]\bigg\rangle_\rho L^{-4/9}  +
\frac{4 \overline{E}_q}{3} \langle q^\dagger q \rangle_\rho
(\langle q^\dagger q \rangle_\rho.
\end{eqnarray}
\end{widetext}

\end{document}